\documentclass[paper,12pt]{JHEP}
\usepackage[centertags]{amsmath}
\usepackage{amsfonts} \usepackage{amssymb} \usepackage{amsthm}
\allowdisplaybreaks[1]
\usepackage{graphicx}
\usepackage{mathrsfs}

\newcommand{\ii}{\mathrm{i}}
\newcommand{\ee}{\mathrm{e}}

\newcommand{\one}{{\rm 1\kern -.9mm l}}

\newcommand{\ft}[2]{{\textstyle\frac{#1}{#2}}}

\newdimen\tableauside\tableauside=1.0ex
\newdimen\tableaurule\tableaurule=0.4pt
\newdimen\tableaustep
\def\phantomhrule#1{\hbox{\vbox to0pt{\hrule height\tableaurule
width#1\vss}}}
\def\phantomvrule#1{\vbox{\hbox to0pt{\vrule width\tableaurule
height#1\hss}}}
\def\sqr{\vbox{%
  \phantomhrule\tableaustep
\hbox{\phantomvrule\tableaustep\kern\tableaustep\phantomvrule\tableaustep}%
  \hbox{\vbox{\phantomhrule\tableauside}\kern-\tableaurule}}}
\def\squares#1{\hbox{\count0=#1\noindent\loop\sqr
  \advance\count0 by-1 \ifnum\count0>0\repeat}}
\def\tableau#1{\vcenter{\offinterlineskip
  \tableaustep=\tableauside\advance\tableaustep by-\tableaurule
  \kern\normallineskip\hbox
    {\kern\normallineskip\vbox
      {\gettableau#1 0 }%
     \kern\normallineskip\kern\tableaurule}%
  \kern\normallineskip\kern\tableaurule}}
\def\gettableau#1 {\ifnum#1=0\let\next=\null\else
  \squares{#1}\let\next=\gettableau\fi\next}
\def\XXint#1#2#3{{\setbox0=\hbox{$#1{#2#3}{\int}$}
     \vcenter{\hbox{$#2#3$}}\kern-.5\wd0}}

\usepackage{url}

\def\be{\begin{equation}}
\def\ee{\end{equation}}
\def\bea{\begin{eqnarray}}
\def\eea{\end{eqnarray}}
\newcommand{\nn}{\nonumber}
\def\ii{{\rm i}}

\title{Fuzzballs in general relativity: a missed opportunity  
}
\author{Lorenzo Pieri$^{1,2}$
\\
\vskip 0.2cm
$^1$ Universit\`a di Roma Tor Vergata and I.N.F.N, Dipartimento di Fisica, 
\\ Via della Ricerca Scientifica, I-00133 Rome, Italy\\
\vskip 0.2cm
$^2$ Centre for Research in String Theory,  School of Physics and Astronomy, \\
Queen Mary University of London, \\Mile End Road,  London, E1 4NS, United Kingdom \\
\vskip 0.2cm
\email{lorenzo.pieri@roma2.infn.it}
}
\abstract{Recently some 4d asymptotically $AdS_2 \times S^2$ regular bubbling solutions written in terms of an arbitrary profile function has appeared in literature. We discuss the possibility of extending these solutions to asymptotically flat spaces, therefore building a model for an horizonless and singularity-free fuzzball with the same asymptotic charges  of the associated black hole, directly in general relativity. A negative conclusion is  reached within an axial-symmetric ansatz inside the IWP family.}


\preprint{ROM2F/2016/09}

\begin{document}
\begin{Huge}

\end{Huge}
\pagebreak

\tableofcontents

\pagebreak

\section{Introduction}

The fuzzball proposal associates to every black hole microstate a regular and horizonless supergravity solution, getting rid of the general relativistic issues of curvature singularity and event horizon \cite{Mathur:2005zp}. 
 
While something is known about fuzzballs of five dimensional black holes, the problem of finding purely regular four-dimensional solutions has been only recently explored by Lunin in  \cite{Lunin:2015hma}, where a class of regular solutions pertaining to the IWP family (after Israel, Wilson and Perjes \cite{Israel:1972vx,Perjes:1971gv})  with asymptotic $AdS_2 \times S^2$  geometry was constructed. The IWP solutions are specified by a complex harmonic function $H=H_1 + i H_2$, and in   \cite{Lunin:2015hma} such harmonic included an arbitrary closed  profile curve $\vec{F}(v)$ with values on $\mathbb{R}^3$. The presence of an arbitrary function has already been important in the past to account the microstates of the two charge system \cite{Lunin:2001jy}: the clearest picture comes from the F1-P system in which the curve can be thought as the oscillation of the string in the non-compact spaces. Analogously, Lunin's solutions can be viewed as microstates associated to the near horizon geometry of a 4d black hole\footnote{Some authors have suggested that, given the double asymptotic of $AdS_2$, these solutions should thought via $AdS_2 /CFT_1$ correspondence as  entangled states living on the product of two copies of the black hole Hilbert space, rather than microstates of a single black hole \cite{Sen:2011cn,Chowdhury:2015gbk}.}, even though no explicit geometric quantization of these geometries has been performed so far (the 2-charge calculation has appeared in \cite{Rychkov:2005ji}).

One can ask whether these bubbling solution can also be viewed as the near horizon geometry of an asymptotically flat BPS fuzzball geometry regular everywhere. In other words, one would like to construct the fuzzballs of 4d black holes.

A great percentage of the fuzzball literature has focused on finding solution regular in five and six dimensions (see \cite{Bena:2007kg} for a review). The main goal is to find the fuzzball of the three charge black hole, the D1-D5-P system, in such a way to have a description of a regular object replacing a black hole with non zero area in the supergravity approximation. At this stage many progress have been made \cite{Bena:2016ypk}, but we still lack a solution sufficiently generic to account  the microscopic entropy.
In this approach, the four dimensional picture comes from the dimensional reduction of a regular four charge geometry in five or six dimensions. The mechanism of regularization involves the presence of a Kaluza-Klein monopole (also known as Euclidean Taub-NUT): this space is free of curvature singularities and the throats that open up near the centers of the monopoles are not infinitely extended, but instead cap smoothly at a finite distance due to the shrinking of a compact dimension. In this picture the four dimensional solution remains generically singular and it is only in higher dimension that a full regularity is achieved.

In these paper we will follow a different direction and explore the possibility of finding regular solutions directly in four dimensions. In this case, if in addition we want to remain in pure general relativity, there are very stringent theorems constraining the occurrence of asymptotically flat regular geometries \cite{Breitenlohner:1987dg,Gibbons:2013tqa}.  One thing that these theorems fails to take into account is the possibility of considering non trivial topologies for the spacetime, in particular the presence of more than one asymptotic region. Indeed it's well known that $AdS_2$ has a disconnected boundary, therefore  a crude extension to flat space should resembles a topologically non trivial wormhole-like geometry.

Let us restate the question: it is possible to build a 4d fuzzball directly in general relativity extending the regular, but asymptotically $AdS_2\times S^2$, 4d bubbling geometries to asymptotically flat? The same question  for asymptotically $AdS_3 \times S^3$ spaces was answered positively, leading to formulation of the fuzzball proposal.

  In the following we will focus on a particularly symmetric choice of the profile function, a circular one, for which we can extract analytic results. What we consider is a combination that we dub {\it circle-NUT}, combining the circular profile of Lunin's with a single-center charged Taub-Nut located along the axis. As we will see, within the restricted axially-symmetric ansatz, justified by the symmetry of the profile, we don't find any asymptotically flat regular solution. 

 The investigation of  IWP geometries as a potential microstate-carrying geometries was inspired by the recent matching of stringy microstates with supergravity fields for a four charge systems of four intersecting D3 branes \cite{Bianchi:2016bgx,Pieri:2016pdt}. 
 In \cite{Bianchi:2016bgx} a precise dictionary between string microstates associated to open string stretched between the branes and the associated back-reacted supergravity fields has been given, by computing string scattering amplitudes of closed string fields with open string fermions living on the branes.  Even though the IWP is the most simple geometry that can be embedded in type IIB string theory D3 brane system, it is natural to go beyond this class and consider a more general class of  type IIB supergravity solutions, describing intersecting D3 branes, that can be written in  terms of eight harmonic functions $H_a(\vec{x})$, $a=1,\dots 8$, on (flat) $\mathbb{R}^3$. These may be associated to the four electric and four magnetic charges in the type IIA description of the system \cite{Bates:2003vx,Balasubramanian:2006gi,Dall'Agata:2010dy}.  
 In addition to non-trivial metric and gauge-field profiles, the BMP solutions involve non-trivial scalar profiles (corresponding to the internal components of the metric and other potentials) that are not considered in the IWP family. We leave it to a future investigation \cite{BMPZ} to study whether regular horizonless asymptotically flat solutions can be found within the type IIB supergravity regime or in order to recover regularity, which in the full string theory is guaranteed by the fact that these are states on which one can safely scatter probes on.

 The plan of the paper is as follows. In Section \ref{IWPsol} we briefly review IWP type solutions characterized by the choice of one complex harmonic function.  We discuss their asymptotic geometry and charges and present some explicit cases ($AdS_2\times S^2$, Kerr-Newman and Taub-NUT). 
 In Section \ref{Fuzzsol} the 4d bubbling solutions \cite{Lunin:2015hma},
depending by an arbitrary close profile function,  are reviewed.

 In Section \ref{RegSol} we set up the regularity conditions that the metric must fulfill. We look for regular horizonless asymptotically flat solutions and a negative conclusion is reached within the simple axially-symmetric anstaz at least. 
   Various appendices contain detailed derivations of some of the technical results presented as well review the dictionary between disk amplitudes and micro-states.

 \section{IWP solutions} 
 \label{IWPsol}
 
 The IWP family \cite{Israel:1972vx,Perjes:1971gv} contains many  different extremal solutions determined by the choice of the complex harmonic function $H$, ranging from regular and geodesically complete to solutions containing genuine curvature singularities. Particularly famous examples are (global) $AdS_2\times S^2$, extremal Kerr-Newman and charged Taub-Nut solutions. Among the black holes like object, it was conjectured long ago, that inside IWP only the Majumdar-Papapetrou family is free of naked singularities \cite{Hartle:1972ya}, but here we hare interested in horizonless solutions.
 
As we will see,  global $AdS_2\times S^2$  space is described by an harmonic with a circular singularity. Unfortunately, the naive extension of these geometries to  asymptotically flat spaces leads to singular spaces, as in  the case of the  Kerr-Newman metric that extends  $AdS_2\times S^2$ towards a flat infinity, but  encountering a curvature singularity at a finite value of the radial coordinate.

In the following we will briefly review the IWP geometries and present some notable examples. 

We consider  Maxwell-Einstein theory with action
\be
S=\int \sqrt{g} \left[  \ft{1}{ 16 \pi G_N} \, R-\ft{1}{4 } \, F^2 \right]
\ee
where we take $G_N=1/(16 \pi)$. The equations of motion reads
 \bea
 R_{MN}-\ft12 g_{MN}\, R &=&   T_{MN}  \nn \\
  d \star_4 F &=& 0
 \eea
 with
 \be
 T_{MN}=-{\delta S\over \delta g^{MN}  }= \ft{1}{2 }  \left( F_{MP}F_{N}{}^P-\ft14 g_{MN} \, F^2 \right)
 \ee
A class of solutions of these equations known as IWP can be written as
\bea
ds^2 &=& -  \left| H  \right|^{-2} \, (dt+w)^2 + \left| H  \right|^2 \,  d \vec x^2 \nn\\
A &=& {2 H_1 \over \left| H  \right|^2  } \, (dt +w)-2 \, \beta  \label{iwp}
\eea

Geometries with the above form are dubbed conformastationary metrics. Here  $H$ is  a complex harmonic function on $\mathbb{R}^3$:
 \be
 H=H_1+\ii H_2 \qquad\quad \quad  \square_3 H= 0
 \ee
  and $w$ and $\beta$ one forms defined as
   \bea
d\omega&=&-2 \star_3[H_2 dH_1-H_1 dH_2] \nn\\
d\beta &=& \star_3  d H_2   \label{wb}
\eea
 Here $ \star_3 $ is the Hodge dual in $\mathbb{R}^3$ computed with the flat metric.  
 
 IWP solutions are electric magnetic duality-invariant, in particular a duality is simply a change in the phase of the complex harmonic, and they can be made non extremal by introducing an auxiliary function in the solution. See \cite{Ortin:2015hya} for further details.
 
  \subsection{ Asymptotic geometry and charges}
  
We will focus on axially symmetric solutions and write the asymptotic geometry in spherical coordinates
\be
d\vec x^2 =    dr^2+r^2 d\theta^2+r^2\, \sin^2\theta\, d\varphi^2   
\ee
 with
 \be
(x_1,x_2,x_3) = (r \,\sin \theta\, \cos\phi, r\, \sin \theta\, \sin\phi,  r\, \cos\theta)  \qquad    r \in [0,\infty) \quad  \theta \in [0, \pi] \quad  \phi \in [0,2 \pi] \nn\\
\ee
 In these variables an axially symmetric  harmonic function  at infinity can be expanded  as an infinite series with coefficients given by the Legendre polynomials $P_{l}(\cos\theta)$. Up to   second order one can write
 \bea
H_1 &= &c_0+\frac{c_1}{r}+\frac{ c_2\,\cos\theta}{r^2}  +\ldots  \nn\\
H_2 & = & d_0+\frac{d_1}{r}+\frac{ d_2\,\cos\theta}{r^2}  +\ldots  \label{h1h2kn}
\eea
 Using (\ref{wb}) one finds for the $w$ and $\beta$ forms  
 \bea
w & =& \left[ 2  \cos \theta 
   (c_0 d_1-c_1
   d_0)+\frac{2 \,\sin ^2\theta 
   (c_2 d_0-c_0
   d_2)}{r}+      \frac{  \sin ^2\theta  ( -c_1 d_2+c_2
   d_1)}{r^2} \right]d\phi  +\ldots \nn \\
\beta & = & \left(d_1\cos \theta
   -\frac{d_2\sin ^2\theta
   }{r} \right)  d\phi+\ldots 
   \eea
   with dots coming from terms omitted in (\ref{h1h2kn}).  The mass $M$,   the electric  and magnetic charges
 $Q$ and $P$ respectively and the angular momentum $J$ of the solution  can be expressed in terms of the coefficients $c_a,d_a$   describing the asymptotic expansion of the complex harmonic function $H$.  The four charges are computed by the integrals
 
 \bea
M &=& \frac{1}{8 \pi}    \int_{  S^2_{\inf}  } \star_4 \,  d  \xi^{(t)} ={c_0 c_1+d_0  d_1\over c_0^2+d_0^2}
     \nn\\
 J &=&- \frac{1}{16 \pi}  \int_{  S^2_{\inf}  }  \star_4 d \xi^{(\varphi)} =c_0 d_2- c_2 d_0  \nn\\
Q &=& \frac{1}{8 \pi}  \int_{  S^2_{\inf}  }  \star_4  F = c_1 + 2  d_0 {c_0 d_1-c_1 d_0  \over c_0^2+d_0^2} \nn\\
P &=& \frac{1}{8 \pi} \int_{  S^2_{\inf}  }   F = d_1  - 2 c_0  {c_0 d_1-c_1 d_0  \over c_0^2+d_0^2} \label{charges}
\eea
The absence of Dirac-Misner strings requires that   $w$ vanishes at infinity, or equivalently
    \bea
&&  c_0 d_1-c_1 d_0=0 \label{cond2}
   \eea

\subsection{Examples}

\subsubsection{$AdS_2 \times S_2$ solution}

To embed $AdS_2 \times S_2$ into the IWP geometry is convenient to use the oblate coordinates defined as
\bea
(x_1,x_2,x_3) &=& (r \,\sin \theta\, \cos\phi, r\, \sin \theta\, \sin\phi,  r\, \cos\theta) \nn\\
&=& \left(  \sqrt{(\rho^2 +L^2) (1-\chi^2) }  \, \cos\phi, \sqrt{(\rho^2 +L^2)(1-\chi^2) }\, \sin\phi,  \rho\,  \chi \right)  \label{change}
\eea
with  $L$ a  constant and
\bea
 && \rho \in (-\infty,\infty) \quad\quad \chi \in [-1, 1] \quad\quad \phi \in [0,2 \pi]
\eea
 Notice that the oblate coordinates cover twice the flat space since the points $(\rho,\chi)$ and $(-\rho,-\chi)$ are identified. 
In the oblate variables the three dimensional flat metric reads 
 \bea
d\vec x^2 
&=&   { \rho^2+L^2 \chi^2 \over \rho^2+L^2}  d\rho^2+  ( \rho^2+L^2 \chi^2  ) {d\chi^2 \over 1-\chi^2}+ (\rho^2+L^2) (1-\chi^2)  d\varphi^2  
\label{dxoblate}
\eea
  The $AdS_2\times S^2$ metric follows by choosing 
\bea
H=\frac{q}{\sqrt{ x_1^2+x_2^2+  ( x_3- \ii \, L)^2 } }=\frac{q}{\rho- \ii \, L  \chi }
\label{ads2}
\eea
leading to
\bea
w &=& \frac{L  \,q^2 \,  (1-\chi^2)   }{ \rho^2 + L^2\, \chi^2 } d \phi \nn\\
\beta &=& \frac{L  \,q \, \rho\, (1-\chi^2)  }{ \rho^2 + L^2\, \chi^2 }   d \phi   \label{wbads2}
\eea
 Plugging (\ref{dxoblate}-\ref{wbads2}) one finds the $AdS_2\times S^2$ metric in global coordinates, that is a geodesically complete space.

\subsubsection{Kerr-Newman solution }

Next we consider the choice
\bea 
H=1+\frac{q}{\rho- \ii \, L  \chi }   \label{hkn}
\eea
leading to 
\bea
w &=& \frac{L  \,q \,( q+2 \rho)\, (1-\chi^2)  }{ \rho^2 + L^2\, \chi^2 }  d \phi\nn\\
\beta &=& \frac{L  \,q \, \rho\, (1-\chi^2) }{ \rho^2 + L^2\, \chi^2 }  d \phi   \label{wbkn}
\eea
The charges of the solution can be read from the large $r$ expansion of $H$. From (\ref{change}) one finds 
\bea
\rho= r - \frac{L^2  (1-\chi^2)}{2 r} +  \ldots \qquad 
\chi=\cos\theta +\frac{L^2 \,\sin^2\theta \, \cos\theta}{
 2  r^2}+ \ldots
\eea
leading to
\bea
H_1 &=& 1+{q\over r}+ {\rm i}\,  {q\,L \, \cos\theta \over r^2}+  \ldots
\eea
 The mass and charges follows from (\ref{charges}) 
 \be
 M=Q=q    \qquad\qquad   J=q\, L  \qquad   \qquad P=0
 \ee
 Moreover plugging (\ref{dxoblate},\ref{hkn},\ref{wbkn}) into (\ref{iwp}) one finds the Kerr-Newman metric.    
In these coordinates the metric has been already analytically extended beyond the horizon (see the appendix B for the usual Boyer-Lindquist set of coordinates), but the resulting metric has a curvature singularity  for $\rho=-q$ and $\chi=0$. 
Indeed:
\bea
R_{\mu \nu} R^{\mu \nu} =\frac{4 q^4}{\left[ (q+\rho)^2 + L^2 \chi^2\right]^4}
\eea

Here we immediately see that adding a simple constant to the geodesically complete $AdS_2\times S^2$ gives a pathological space.

\subsection{Charged Taub-NUT and Reissner-Nordstom solutions}

Finally we consider the choice
\be
H  =1+\frac{b_1+i b_2}{ \sqrt{ \rho^2+ L^2   (1-\chi^2) }} \label{hnut}
\ee
leading to
\be
w= 2\,\beta=- \frac{2 \,b_2  \, \chi}{\sqrt{  \rho^2+ L^2   (1-\chi^2)   }} d \phi  \label{wbnut}
\ee
Plugging (\ref{dxoblate}, \ref{hnut}, \ref{wbnut}) into (\ref{iwp}) one finds the charged Taub-Nut metric \cite{Newman:1963yy} with
charges given by (\ref{charges}),
\be
M=Q=b_1  \qquad\qquad   P=-b_2  \qquad\qquad   J=0
\ee 
This space is regular at the potentially troublesome point $\rho=0, \, \chi=1 $.   Indeed one can see that there is no curvature singularity anywhere by evaluating the Riemann tensor in an orthonormal frame. Still, the metric has a Dirac-Misner string singularity, that takes the name from the similar configuration of the Dirac string for magnetic monopoles. The string appears  since there is no gauge choice in which $w$ is  vanishing at both $\chi=1$ and $\chi=-1$ \footnote{It's still possible to remove the Dirac-Misner string, but the price to pay is to compactify the time direction, as shown by Misner \cite{Misner:1963fr}. Anyway further problems remains, like the presence of CTCs. We notice that a recent analysis of the geodesic motion in Taub-NUT points out that the unphysical regions can actually be harmless \cite{Clement:2015cxa}. If this view is correct, some of the NUTty fuzzball solutions found in this work can be seen from a new perspective. }. 
String singularities will be recurrent in the following and they will frequently exchange with curvature singularities for certain choice of the parameters. Indeed the relation between these kind of singularities has been explored in the literature, for instance to space-time with string singularities one can associate an entropy, much like for space with horizons \cite{Hawking:1998ct,Mann:1999pc}.

 Finally, the extremal Reissner-Nordstom (ERN) black hole metric corresponds to taking $b_2=0$. In this limit the string singularity disappear, but on the contrary the curvature singularity at $\rho=-b_1$, $\chi=1$ shows up. 
 
 Notice that in order to see the singular point one must analytically continue the square root in the oblate coordinate system. This is a bit confusing, indeed while the $\lbrace \rho, \chi \rbrace $ are perfectly suited to analytically extend a space singular on an ring, they are quite awkward   coordinates to deal with a point-like singularity. The ERN black hole harmonic is usually written in isotropic coordinates as:
 
 \bea
 H=1+{b_1 \over |x| } =1+{b_1 \over r }
 \eea
 
 with $r \in (0,\infty) $ and $b_1=M=Q$. Since the location of the horizon ($r=0$) is a perfectly regular point, this coordinate system is geodesically incomplete and one must consider negative values of  $r $. Given that the relation \eqref{change} between  spherical coordinates $\lbrace r, \theta \rbrace $ and  oblate coordinates $\lbrace \rho, \chi \rbrace $ implies $r = \pm \sqrt{\rho^2+ L^2   (1-\chi^2)}  $ we see that starting for big values of $\rho$, that is far from the center of the hole, we enter the horizon only if both $\rho=0$ and $\chi= \pm 1$ are satisfied. A negative values of rho does not therefore indicate a priori that we have entered into the horizon.
 
\section{ Fuzzball solutions}
\label{Fuzzsol}

 In the spirit of the fuzzball proposal one can ask whether a family solutions of IWP type with no horizon or singularities  exist. In fact a family of asymptotically $AdS_2\times S^2$ geometries of IWP type regular everywhere has been recently constructed  in \cite{Lunin:2015hma}\footnote{In the same paper there is a section on asymptotically flat solutions. Unfortunately a misprint in the harmonic function invalidates the analysis.}, where the solutions are specified 
 by a profile function $\vec{F}(v)$ with values on $\mathbb{R}^3$ describing the singularities of the harmonic function $H$.
 In particular, the global $AdS_2\times S^2$  space is describe by a circular profile, while slight deformations of the profile lead to bubbling of the $AdS_2\times S^2$  geometry.  Similar configurations have been also considered in \cite{Park:2015gka} in the context of exotic branes and non geometric black hole microstates.
 
 Despite the singularities of $H$ along the profile, the resulting geometries where proven to be regular everywhere. The mechanism of regularization is different from  previous cases in literature: in  $AdS_5\times S^5$ the geometry terminates before reaching a singular point, since an $S^3$ of the geometry collapses to zero size \cite{Lin:2004nb}; similarly in $AdS_3\times S^3$ a sphere collapses and the space ends smoothly in the shape of a KK monopole \cite{Lunin:2002iz}. In  $AdS_2\times S^2$ the regularization comes from gluing different $\mathbb{R}^3$ 
through one or more branch
cuts, as could be expected by the fact that the boundary of $AdS_2$ is disconnected, while the one of $\mathbb{R}^3$ it is not.
 
 We consider IWP geometries characterized by a harmonic function $H$ of the form

\bea
 H= h_{reg} 
 +    \int_0^{2\pi} \, {dv \over 2\pi } \frac{1}{|\vec{x}-\vec{F}|}  \sqrt{1+ \frac{(\vec{x}-\vec{F})\vec{A}}{|\vec{x}-\vec{F}|^2}}    \label{Hhx}
\eea
with  $\vec{A}(v) $ a complex periodic vector with values on $\mathbb{C}^3$ satisfying $\vec{A}\cdot \vec{A}=\dot{\vec{F}}\cdot\vec{A}=0$. 
The condition $\vec{A}\cdot \vec{A}=0$ guarantees that $H$ is harmonic. The condition $\dot{\vec{F}}\cdot\vec{A} = 0$ ensures that the metric is regular near the curve, as we will show. The remaining $h_{reg}$ must be a piece regular on the location of the curve; in the case in which is regular everywhere, the conditions to impose for the non vanishing of $H$ have been explicitly written in \cite{Lunin:2015hma}.

To check regularity near the location of the profile, let us consider a point $\vec x$ near the curve, and choose local coordinates cylindrical coordinates such that  $\vec{F}(v)\approx (0,0,v)$ near $v=0$, that is $\vec{F}(v)$ is along the axis of the cylinder,    
and $\vec x=( \kappa \cos\zeta,   \kappa \sin\zeta, 0)$ is a vector in the plane perpendicular to the curve.  The condition  $\dot{\vec{F}}\cdot\vec{A} = 0$  requires then 
  $\vec{A} =a(1,\ii,0)$. In this near-curve limit $\kappa \rightarrow 0$ the profile appears as a infinite straight line in the $z$ direction and the integral in \eqref{Hhx} approximates to:
  \bea
H\approx     \int_{-\infty}^{\infty} \, {du \over 2\pi \sqrt{ \kappa }} \frac{\sqrt{a }e^{i \zeta/2}}{1+u^2} =    \sqrt{a  }\frac{e^{i \zeta/2}}{2\sqrt{ \kappa } } 
  \eea
having defined the dimensionless variable $u=v/\kappa$. On the other hand one finds for the $w$-function
  \bea
  \omega \approx  \frac{ a \, d z }{ 4\kappa}
\label{nearcurve}
\eea
  leading to a near flat metric  
 \bea
ds^2 &\approx & - \frac{4\kappa }{a} \left(dt+ \frac{a \, dz }{ 4\kappa }\right)^2 +\frac{a}{ 4\kappa}  (d\kappa^2+\kappa^2 d\zeta^2+   dz^2)  \\ \nn
&\approx & - 2\,dt dz  +  d \hat \kappa^2+\frac{\hat \kappa^2}{4} \, d \zeta^2
\eea
with $\hat \kappa=\sqrt{a \kappa}$. To avoid a conical singularity the angle must live in the range $\zeta \in [0,4 \pi)$. Indeed when we make a full $2 \pi$ turn around the profile, we must cross the branch cut surface with endpoint on the profile, therefore going from one $\mathbb{R}^3$ to the other; it is only the double cover of the original $\mathbb{R}^3$ that is  regular. This situation can be compared to the difference between $AdS$ in the Poincar\'e patch and in global coordinates.

\subsection{Circular profile}

  It's interesting to consider a circular profile 
  
  \bea
  \vec F &=&L(\cos v,\sin v,0) \nn\\
  \vec A &=& 2L(\cos v,\sin v,\ii)  
  \eea
  
  in which these  calculations can be done explicitly and the integral \eqref{Hhx} can be computed analytically, resulting in \eqref{ads2}. With this choice:
  
\bea
 \left| H  \right|^2   & \approx &  {  q^2 \over \rho^2+L^2 \chi^2 }   \qquad     w \approx  {\, L\,q^2\,    \over \rho^2+L^2 \chi^2 }  
\eea
and the metric becomes flat
\bea
 ds^2 \approx   - 2\, L  \, dt \,d\varphi \, 
  +q^2 \left(  {   d\rho^2 \over  L^2 }           +    \, d\vartheta^2 \right)
\eea
  Notice that here is crucial that the domain of $\rho $ is extended to negative values otherwise we would have only half of the flat space. 
 
\section{Circle-NUT solution: Regularity Analysis}
\label{RegSol}

The solutions described in the previous sections can be used as building blocks of  more complicated solutions involving the superpositions of multi-center Kerr-Newman black holes  and charged Taub-NUTs. 
In the following we will restrict ourselves to axially symmetric solutions, therefore considering only solutions centered on the axis of symmetry.
This choice has two main motivations: 1) The circle-source can be seen as a very symmetric choice of the generic curve profile. In the D1-D5 fuzzball and D1-D5-P system different profile functions were related to states in the holographic dual $1+1$ CFT \cite{Lunin:2001jy}, in particular the Fourier decomposition in frequencies of the curve are linked to twisted states in the CFT (for a more precise statement of the correspondence and examples  see  \cite{Kanitscheider:2006zf,Kanitscheider:2007wq,Giusto:2004id}). The most simple CFT states, the untwisted ones, are eigenstates of the R-charge operator in the CFT, translating on the gravity side in the independence of the solution from the angle associated to the R-charge operator. Even though in the 4d case we haven't the same control on the dual CFT, is plausible that the circular profile should be again associated to an angle-independent configuration. 2) It drastically simplifies the treatment with respect to a less symmetric ansatz.

Therefore we consider the superposition of a circular-profile bubbling solution  and a charged Nut with center on the axis perpendicular to the circle singularity \footnote{Besides the constant term, there is another oblate spheroidal harmonic that is constant at both infinities: $Arctan(\rho)$. One can show that its presence leads to sub-leading $log$ divergences in $\omega$ on the profile. The leading divergences are removed as in the previous section, but the sub-leading terms remain, meaning that the circular profile is genuinely singular.}

 \bea
H= a_1+{\rm i} a_2 +  \  \frac{b_1+{\rm i} b_2 }{  \sqrt{ (\rho^2 +L^2 ) (1-\chi^2) + (x_0 - \rho \chi)^2 }} +     \frac{q (\rho+{\rm i}\,  L \,\chi) }{\rho^2+ L^2\chi^2 }
\label{circlenuth}
\eea

for the sake of simplicity let's focus on the case $x_0=0$, the general case can be found in the appendix. This choice of the harmonic leads to:

\bea
w &=& c+ \frac{q \left(  \rho^2 +L^2   \right) \left(2 a_2 \chi L  +q \right)+2 a_1  q  L^2 \rho 
   \left(1-\chi ^2\right)}{L \left(  \rho^2+   L^2 \chi ^2\right)}+   \frac{2  \chi  \rho \left(a_2 b_1-a_1 b_2\right)}{\sqrt{  \rho^2+ L^2   (1-\chi^2)   }}  \nn\\
&&   +\frac{2 \left(b_2  q \chi  L^3\left(1-\chi
   ^2\right)+b_1 q  \rho  \left( \rho ^2+   L^2\right)\right)}{L
  \rho \, \left(  \rho ^2+ L^2 \chi ^2\right) \sqrt{  \rho^2+ L^2   (1-\chi^2)   }}  \nn\\
  \beta &=&  \frac{  \rho\, q \, L 
   \left(1-\chi ^2\right)}{ \left(  \rho^2+   L^2 \chi ^2\right)}- \frac{b_2\,  \chi \rho }{\sqrt{  \rho^2+ L^2   (1-\chi^2)   }}     \label{wbcnut}
   \eea
with $c$ a constant. 

The charges can be read from the large $\rho$ expansion 
\bea
H_1 &=& a_1+{b_1+q\over  r}+O(r^{-3}) \nn\\
H_2 &=& a_2+{b_2 \over r}+{L q\, \chi \over r^2}+O(r^{-3}) 
\eea
leading to 
\bea
 M &=&  {a_1 (b_1+q)+a_2  b_2\over a_1^2+a_2^2} 
     \nn\\
 J &=& a_1 \, L\, q   \nn\\
Q &=&  b_1+q +  { 2 \,q \,a_1\, a_2   \over a_1^2+a_2^2} \nn\\
P &=&   b_2  -   { 2\, q\, a_1  a_2 \over a_1^2+a_2^2} \label{charges}
 \eea

The presence of the square root in the metric requires great care. While these coordinates are adapted to the circle-harmonic, the analytic extension of the geometry implies that we need to consider both positive and negative values for the square root, as already seen with Taub-NUT. In particular while the pure circle-harmonic geometry connect two different asymptotic $\mathbb{R}^3$ regions \cite{Lunin:2015hma}, the circle-point geometry connect four different $\mathbb{R}^3$ (and in general $2^{n+1}$ $\mathbb{R}^3$ regions for n Taub-NUT centers), since we have two possible signs for the square root, going from one to the other by crossing the point  $\rho=0, \chi=\pm 1$.

\subsection{Regularity conditions}

As we will see, the quest for a regular solution inside IWP amount to a delicate balance between 
removing Dirac-Misner strings and avoiding curvature singularities. Even though it's possible to find such configurations, the additional request of asymptotic flatness seems to be to demanding to be accommodate inside the axially-symmetric IWP ansatz.

 Curvature singularities can arise at the boundaries of the space or on interior points where  some of the eigenvalues of the metric vanish or diverge.   Writing the IWP metric in the form
\bea
ds^2 &=& -  \left| H  \right|^{-2} \, (dt+w)^2 + \left| H  \right|^2 \,   
\left(   L_\rho^2\,  d\rho^2+  L_\chi^2 \, d\chi^2+   L_\varphi^2\,  d\varphi^2  \right)
\eea
with
\be
 L_\rho^2=  { \rho^2+L^2 \chi^2 \over \rho^2+L^2}   \qquad      L_\chi^2=  { \rho^2+L^2 \chi^2  \over 1-\chi^2}   \qquad   
 L_\varphi^2=  (\rho^2+L^2) (1-\chi^2) 
\ee
one finds for the eigenvalues
\bea
\lambda_1 &=& \left| H  \right|^2 \,  \,   L_\chi^2  \qquad      \lambda_2= \left| H  \right|^2 \,  \,   L_\rho^2  \\
\lambda_{3,4} &=& -{1\over 2 \left| H  \right|^2  } \, \left( 1+w^2-L^2_\varphi \,\left| H  \right|^4  \pm \sqrt{   4\, L_\varphi^2  \,\left| H  \right|^4 +
 ( 1 +w^2-L^2_\varphi \,\left| H  \right|^4   )^2     }  \right)  \nn
\eea
 For $L>0$, the eigenvalues can vanish or diverge when one of the following conditions hold
 \begin{itemize}
 
  \item{ {\bf A:}  $ \rho=-\infty$ and $ \rho=\infty$}
 
 \item{ {\bf B:}  $ \chi=\pm 1 $ }
 
 \item{ {\bf C:}  $\left| H  \right|^2 =\infty $  or   $w=\infty$  }
 
  \item{  {\bf D:}  $\left| H  \right|^2 =0 $ }
  
   \end{itemize}

 \subsection*{ Case A: $ \rho=\pm\infty$}
 
 The metric is asymptotically AdS with the choice $a_1=a_2=0$, while it's asymptotically flat if at least one of the two constants is different from zero.
 
 \subsection*{ Case B: $ \chi=\pm 1$}

Around these points, $w=0$ and the metric becomes
\bea
ds^2 &=& -  \left| H  \right|^{-2} \,  dt^2 + \left| H  \right|^2 \,   
\left[  d\rho^2+   (\rho^2+L^2 ) d\Omega_2  \right] 
\eea
 This metric can fail to be regular only at point where $|H|$ either vanishes or diverges, so regularity boils down to check conditions C and D.

  \subsection*{ Case C: $ |H|=\infty$}

The harmonic function diverges at two points:
\bea
(\rho,\chi) &=&(0,\pm 1)  \qquad \Rightarrow    H\approx H_{TN} \nn\\
(\rho,\chi) &=& (0,0)  \qquad \Rightarrow    H\approx H_{AdS_2\times S^2} 
\eea
 leading in both cases to regular geometries near $\rho=0$. 
 
 \subsection*{ Case D: $ |H|=0$}
    
    The conditions $H_1=H_2=0$ boils down to 
      \bea
&&  b_2 a_1  -b_1 \, a_2 +        \frac{ b_2 \,q\, \rho- q  b_1 \,  L \,\chi   }{\rho^2+ L^2\chi^2 }=0 \nn\\
&&  q \, \rho \, a_2- L\, \chi\, a_1  +        \frac{ b_2 \,q\, \rho-q  b_1 \,  L \,\chi  }{ \sqrt{ \rho^2+ L^2(1-\chi^2)} }=0 
\label{h1h2}
\eea
 The solution is regular if  it exists a choice of parameters  such that these two equations have no solution. For some choice of the parameters the solution is particularly simple to analyze; we have already seen that it reduces to $AdS_2 \times S^2$, Kerr-Newmann and Taub-NUT.

Some other notable solutions are:
 
 \begin{itemize}
 \item $a_1=b_1=b_2=0$
 
  \bea
H= {\rm i} a_2 +  \frac{q (\rho+{\rm i}\,  L \,\chi) }{\rho^2+ L^2\chi^2 }
\eea
This space is asymptotically flat, but with a Dirac-Misner sting. The space is completely regular in the curvature for $\left |\frac{q}{L a_2} \right | >1$, indeed since the real part is zero only for $\rho=0$, the complex part requires $\chi=-\frac{q}{L a_2}$.

 \item $a_1=a_2=0$

The first equation of \eqref{h1h2} implies that $  \rho=\frac{  b_1 \,  L \,\chi}{b_2 }   $. Substituting $\chi$ into $H_2$  gives two solutions: 

\bea
\rho= \pm \frac{b_1 \, L \, q \,} {\sqrt{(b_1^2 + b_2^2)^2 + q^2 (b_2^2 - b_1^2)}} 
\eea

The geometry is free of curvature singularities for $(b_1^2 + b_2^2)^2 + q^2 (b_2^2 - b_1^2)<0$. Unfortunately it turns out that it has Dirac-Misner strings. 
If the position of the center is displaced from the origin, is possible to cancel the Dirac-Misner string for the choice $x_0=\frac{b_1L}{b_2}$ from \eqref{dmx0}, but then a curvature singularities appears, since there is no value of the parameters for which the harmonic is always different from zero.

For  generic non zero choice of the parameters one can compute the value of $w$ for $\chi= \pm 1$, for both branches of the square root:

\bea
w|_{\chi=1}=\frac{1}{L} [\mp 2 a_1 b_2 L + c L \pm 2 b_1 q + q^2 + 2 a_2 L (\pm b_1 + q)]
\eea

\bea
w|_{\chi=-1}=\frac{1}{L} [\pm 2 a_1 b_2 L + c L \pm 2 b_1 q + q^2 - 2 a_2 L (\pm b_1 + q)]
\eea

where $\pm$ comes from the two branches. Assuming $q \neq 0$, the condition $w|_{\chi=1}= w|_{\chi=-1}=0$ forces us in two possibilities: $a_1=a_2=b_1=0$ or $a_2=b_1=b_2=0$. The second possibility is simply Kerr-Newmann, and the first possibilities has always curvature singularities: $H_1$ is zero for $\rho=0$, meaning that $H_2$ is zero for $\chi=\pm \frac{q}{\sqrt{b_2^2+q^2}}$, that has always solution.

\end{itemize}

  \section{ Conclusions} 
 
In this work we have attempted to find a completely  regular asymptotically flat solution in general relativity carrying the same charges of a black hole and that could account for the microscopic entropy due to the degeneracy of the microstates, by  extending a previously know regular asymptotically $AdS_2 \times S^2$  solution. The whole analysis was inspired by the construction of the stringy system of D3 branes, in which the microstates are the different configuration of open strings stretched between the branes, and the matching of these microstates with the SUGRA excitations. The IWP ansatz surveyed in this work is the simplest possible solution contained in the more general SUGRA bound state of D3 branes, since in particular it contains no other fields than pure general relativity.

The analysis carried in this work seems to suggest that there is a conflict between regularity and asymptotic flatness. There are many different ways to interpret these indications, so we list them in increasing order of divergence from the strategy adopted in this paper, delineating possible future directions.

\begin{itemize}
\item The axially symmetric ansatz is not enough. One can argue that with a more general dependence of all the coordinates, is possible to find a regular asymptotically flat solution.
While this observation is reasonable, the geometry associated to a circle profile should be highly more symmetric that any other possible closed profile and previous work in the D1-D5 system has shown that very simple states in the CFT correspond in general to pretty simple geometries. 

\item The IWP ansatz is not enough. It's too optimistic to find a regular solution in general relativity, indeed the internal structure of string theory is crucial for the microstates counting and this suggest that the scalar fields obtained from the compactification of this internal space must play a role. 
In this view, the most simple candidate for a regular solution is the SWIP solution \cite{LozanoTellechea:1999my}, that is again contained in the SUGRA geometries associated to the D3 bound state. The SWIP solution is still supersymmetric and extremal, and contain one single complex scalar field.

 More generally, the four dimensional solution of the systems of 4 D3s reduced on $T^6$ can be seen as  ${\cal N}=2$ truncation of  ${\cal N}=8$ supergravity involving the gravity multiplet plus three vector multiplets \cite{BMPZ}, and it can contain up to 3 complex scalar fields associated to internal structure of the tori.

\item Regularity can be achieved only in higher dimension, not in 4d. Is the compactification to 4d that creates artificially a singularity, but the true theory is completely regular and the information paradox never arises. One must focus on 5d and 6d models to solve the problem. 

Notice that a regular solution in higher dimension, can be read in 4d language as if a particular combination of the metric and the scalars is regular, even though the metric and the scalar alone have issues. Therefore in some sense this possibility overlaps with the previous point in the list.

\item It is not possible to find regular solution directly in supergravity at all. The success of the D1-D5 was a miracle of 6d physics, but 4d  physics is different. Even though these solutions appear to be singular in the SUGRA limit there is nothing wrong with them, since branes and strings are  basic objects string theory, so they are well defined in the full theory.

\end{itemize}

\section*{Acknowledgements}

A special thanks to M. Bianchi, J. F. Morales, O. Lunin and N. Zinnato for reading the manuscript and collaborating to the early stage of the project.
We would like also like to thank I. Bena, E. Moscato, R. Panerai, R. Russo, K. Skenderis for useful conversations. This work is partially supported by the MIUR PRIN Contract 2015MP2CX4 "Non-perturbative Aspects Of Gauge Theories And Strings".

\begin{appendix}

 \section{  Intersecting D3-branes } 
 
 In this section we briefly review a  class of  type IIB supergravity solutions describing intersection of D3 branes, which reduce to IWP solutions for a particular choice, that can be written  in  terms of eight harmonic functions 
  ($a=1,\dots 8$, $I=1,2,3$) 
   \be
   H_a=\{ V, L_I, K_I, M \} \label{harmh}
   \ee   
  on (flat) $\mathbb{R}^3$ associated to the four electric and four magnetic charges in the type IIA description of the system. 
   These functions are conveniently combined into  
  \bea
 P_I &=& {K_I \over V} \nn\\
 Z_I &=& L_I +{ |\epsilon_{IJK}|\over 2} {K_J K_K\over V} \nn\\
 \mu &=& { M\over 2} +{L_I K_I \over 2\, V}+{ |\epsilon_{IJK}| \over 6} \, {K_I K_J K_K \over V^2}  \label{mudef}
 \eea
 Here
  $\epsilon_{IJK}$ characterize the triple  intersections  between two cycles on $T^6$.
In these variables the supergravity solutions can be written as
      \bea
  ds^2  &=& -   e^{2U}( dt+w)^2 +e^{-2U} \,  \sum_{i=1}^3 dx_i^2 +   \sum_{I=1}^3   \left[  { d y_I^2 \over  V e^{2U} Z_I }  + V e^{2U} Z_I  \,  \tilde e_I^2  \right] \nn\\
  C_4 &=& \alpha_0  \wedge \tilde e_1\wedge \tilde e_2\wedge \tilde e_3+ \beta_0  \wedge dy_1\wedge  dy_2\wedge  dy_3 \nn\\ 
  && +\ft12 \epsilon_{IJK}\,   \left( \alpha_I  \wedge  dy_I \wedge \tilde e_J   \wedge \tilde e_K + \beta_I  \wedge  \tilde e_I \wedge dy_J   \wedge dy_K  \right)   \label{d34}
 \eea
 with  
  \bea
  e^{-4U} &=&  Z_1 Z_2 Z_3 V-\mu^2  V^2 \qquad ~~~~~~~~~~ \nn \\
   b_I &=&     P_I-\frac{\mu}{Z_I}    \qquad \qquad ~~~~~~~~~~~~~~~~~ \tilde e_I =   d \tilde y_I   -  b_I\, dy_I   \nn\\
 \alpha_0 &=& A-\mu\, V^2\, e^{4U}\, (dt+w) \qquad ~~~~~
\alpha_I =   -\frac{(dt+w)}{Z_I} + b_I\, A+w_I   \nn\\
\beta_0&=& -v_0 +{e^{-4U}\over V^2 Z_1 Z_2 Z_3}(dt+w)-b_I \,v_I+ b_1 \,b_2\, b_3 \, A +{|\epsilon_{IJK}|\over 2}\,b_I \,b_J\, w_K  \nn\\
\beta_I &=& -v_I + {|\epsilon_{IJK}|\over 2}\, \left(  {\mu\over Z_J Z_K}\, (dt+w)+b_J \, b_K\, A+2 b_J\, w_K  \right) 
 \eea
   and
  \bea
 {*_3}dA &=& d V     \qquad {*_3}dw_I = -d (K_I)  \qquad    {*_3}dv_0 =dM  \qquad    {*_3}dv_I =dL_I\nn\\
   \qquad  {*_3}dw  &=&  V d \mu-\mu dV-V Z_I dP_I  
  \eea
 These solutions have been recently  interpreted as micro state of the D3-brane system \cite{Bianchi:2016bgx}. More precisely, the general solution can be thought as the superposition of three type of solutions  
   \begin{itemize}
   \item{{\bf L}: Solutions with $ K_I=M_I=0$    }
     \item{{\bf K}: Solutions with $ K_I=K_I(x)$ and $M=\sum_{I=1}^3 K_I(x)$.      }
       \item{{\bf M}: Solutions with $ K_I=K_I(x)$ and $M=-\sum_{I=1}^3 K_I(x)$.      }
   \end{itemize}
   The three classes are generated by disks with one, two and four different boundaries respectively. At infinity the three contributions (deviation from flat space) fall off as $r^{-1}$,  $r^{-2}$, $r^{-3}$ respectively, so $K$ and $M$ functions represent higher multipole micro state excitations of the black hole. 

   The IWP solutions correspond to solutions with a trivial internal square metric 
  \bea
V e^{2U} Z =1 \qquad  b_I=0 
\label{nointernal}
\eea
  These equations can be solved in terms of two harmonic functions $H_1$ and $H_2$ via the identifications
\bea
V &=& L_I=H_2  \nn\\
 M& =& -K_I=- H_1
\eea

  \section{Kerr-Newman solution}
\label{KN}

In the Boyer-Lindquist $\lbrace t,  r,\theta,\phi \rbrace$ coordinates
\footnote{ $
 t \in (-\infty,+\infty) \quad\quad  r \in [0,+\infty) \quad\quad \theta \in [0, \pi] \quad\quad \phi \in [0,2 \pi]$. } them metric and gauge field describing the Kerr-Newman solution can be written as
   \bea
 d s^{2}&=& -\left(1-\frac{2Mr}{ R^2}+\frac{(Q^2+P^2)}{R^2}\right) d t^{2}-   \frac{2a  \sin^{2}\theta}{ R^2}   \left[ 2\,M\,  r - (Q^2+P^2)\right]d t\, d\varphi  \nn  \\
&&+\frac{R^2}{\Delta}d  r^{2}+R^2 d\theta^{2}+\frac{sin^2\theta}{ R^2}\left[ (  r^2+a^2)^2-a^2 \Delta sin^2\theta\right] d\phi^{2} \nn\\
A &=& -{4 \over  R^2}  ( Q\,  r+P\, a\, \cos\theta)\,  dt+\frac{4}{R^4} ( Q\, a \, r\, \sin^2\theta+P\, ( r^2+a^2) \cos\theta )\, d\phi
\eea
where
    \bea
R^2( r,\theta)= r^2+a^2 cos^2(\theta) \qquad \Delta( r)= r^2-2M\,  r+a^2+(Q^2+P^2)
\eea
 The parameter $M$ represents the mass, $Q$ and $P$ the electric and magnetic charges and $J=Ma$ the angular momentum. 
   In the case
  \be
  M^2\geq Q^2+P^2+a^2 
  \ee
 the solutions represents a black hole with horizons at\footnote{The locations of the horizons are given by the vanishing of the norm of the normal vector to $r=const.$ hypersurfaces, that is $g^{rr}=0$.}
 \be
  r_{\pm}=M \pm \sqrt{M^2-a^2-Q^2-P^2}
 \ee
   and a curvature singularity at $ r=0$. 
  The area of the black hole horizon is $A=4\pi ( r_+^2+a^2)$.
 The leading asymptotic in the large $r$ expansion is given by: 
\bea
 d s^{2}=-\left(1-\frac{2M}{ r}+o(r^{-2})\right) d t^{2}-\left(4\frac{J\sin^{2}\theta}{ r}+o(r^{-2})\right)d t\, d\varphi  \nn  \\
+\left(1+\frac{2M}{r}+o(r^{-2})\right) d r^2 + r^2 \left(d \theta^2+\sin^2\theta d\varphi^{2}\right)
\eea

\section{Generic Circle-NUT solution}

          \bea
H= a_1+{\rm i} a_2 +  \  \frac{b_1+{\rm i} b_2 }{  \sqrt{ (\rho^2 +L^2 ) (1-\chi^2) + (x_0 - \rho \chi)^2 }} +     \frac{q (\rho+{\rm i}\,  L \,\chi) }{\rho^2+ L^2\chi^2 }
\label{circlenuth}
\eea
leading to 
\bea
w &=&  c - \frac{2 a_1 q \rho}{L}- 
  2 (L^2 + \rho^2) \bigg \lbrace     \frac{-a_1 q \rho}{L \rho^2 + L^3 \chi^2} - \frac{
     q^2 + 2 a_2 L q \chi}{
     2 L \rho^2 + 
      2 L^3 \chi^2} -  \nn \\
      &-& \frac{(L^2 + 
          x_0^2)^{-1}}{ (L^4 + 4 \rho^4 + 
          L^2 (x_0^2 + 4 \rho^2)} \bigg [  \sqrt{(\rho^2 + 
            L^2 ) (1 - \chi^2) + 
         \frac{C^2}{\rho^2} } \bigg (  \frac{
          b_1 L q \rho (L^2 + x_0^2 + 2 \rho^2 + 
             2 x_0 \rho \chi)}{\rho^2 + 
           L^2 \chi^2} + \nn \\
           &+&  \frac{b_2 a_1 (L^6 + 4 x_0^2 \rho^4 + 
                  2 L^4 D + 
                  L^2 D^2) C + 
               q \rho \big ( 2 x_0 \rho^2 (x_0^2 - \rho^2) +  
                  L^2 (L^2 C+x_0^3 + x_0^2 \rho \chi - 
                    2 \rho^3 \chi) \big )}{(L^2 + \rho^2) \rho^2 \
((1 + L^2/\rho^2 ) (1 - \chi^2) + 
               C^2/\rho^2 )}  \bigg )  \bigg ]  + \nn \\
   &+&                 \sqrt{(\rho^2 + 
            L^2 ) (1 - \chi^2) + 
         \frac{C^2}{\rho^2} } \bigg [ 
          \frac{- b_2 q (-2 x_0 \rho^3 + L^2 x_0^2 \chi +         L^2 (L^2 + 2 \rho^2) \chi}{\rho^2 +             
            L^2 \chi^2} + \nn \\
            &+& \frac{(L^2 + \
\rho^2)^{-1}  \rho^{-2}}{  ((1 + L^2/\rho^2 ) (1 - \chi^2)+              C^2/\rho^2 )}
    \bigg ( b_1 a_2 (L^6 + 4 x_0^2 \rho^4 + 
                  2 L^4 D + 
                  L^2 D^2) C + \nn \\
                  &+&    L q (L^4 (-\rho + x_0 \chi) + 
                  L^2 (-3 \rho^3 + x_0^3 \chi + 
                    2 x_0 \rho^2 \chi) + \nn \\
                    &+&  \rho (x_0^4 - 
                    x_0^2 \rho^2 - 2 \rho^4 + 
                    2 x_0 \rho^3 \chi)) \bigg) \bigg ]  \bigg ( (L^2 + 
          x_0^2) (L^4 + 4 \rho^4 + L^2 (x_0^2 + 4 \rho^2))  \bigg)^{-1}      \bigg \rbrace   \nn\\
  \beta &=&  (-1+\chi^2) \bigg( -\frac{L q \rho}{\rho^2+L^2 \chi^2}- \frac{b_2 (x_0-\rho \chi)}{(1-\chi^2) \sqrt{ (\rho^2 +L^2 ) (1-\chi^2) + (x_0 - \rho \chi)^2 }} \bigg \rbrace    
  \label{wbcnut}
   \eea
with $c$ a constant, $x_0$ the position of the center and we have defined $C=(x_0 - \rho \chi)$ and $D=(x_0^2 + 2 \rho^2)$.

The absence of Dirac-Misner strings  requires that $w$ vanishes at $\chi=\pm 1$ leading to
 \be
 a_1=a_2=0 \qquad  c=- { q^2\over L} \qquad x_0=\frac{b_1 L}{b_2}
 \label{dmx0}
 \ee

 \end{appendix}


\providecommand{\href}[2]{#2}\begingroup\raggedright\endgroup

\end{document}